\documentclass[letter,oldversion]{aa}
\usepackage{natbib}
\usepackage{graphicx}
\usepackage{txfonts}

\begin{document}
\title{HD 17156b: A Transiting Planet with a 21.2 Day Period and an Eccentric Orbit}
\titlerunning{HD~17156 : a new planetary transit}
\authorrunning{M. Barbieri et al.}

\author{M. Barbieri
        \inst{1}
        \and
        R. Alonso
        \inst{1}
        \and
        G. Laughlin
        \inst{2}
        \and
        J. M. Almenara
        \inst{3}
        \and
        R. Bissinger
        \inst{4}
        \and
        D. Davies
        \inst{5}
        \and
        D. Gasparri
        \inst{6}
        \and
        E. Guido
        \inst{7}
        \and
        C. Lopresti
        \inst{8}
        \and
        F. Manzini
        \inst{9}
        \and
        G. Sostero
        \inst{7}
        }

   \offprints{M. Barbieri}

   \institute{
             LAM, Traverse du Siphon, BP 8, Les Trois Lucs,
             13376 Marseille Cedex 12, France\\
             \email{mauro.barbieri@oamp.fr,roi.alonso@oamp.fr}
         \and
              University of California Observatories, University of California at Santa Cruz
              Santa Cruz, CA 95064, USA\\
              \email{laugh@ucolick.org}
         \and
              Instituto de Astrof\'\i sica de Canarias, 
              C/V\'\i a L\'actea s/n, E-38200, La Laguna, Spain
         \and
              Racoon Run Observatory,
              Pleasanton, CA, USA 
         \and
              23819 Ladeene Avenue, Torrance, CA 90505, USA
         \and
             Universit\`a di Bologna, Dipartimento di Astronomia,
             Via Ranzani 1, 40127 Bologna, Italy 
         \and
             Associazione Friulana di Astronomia e Meteorologia, 
             Piazza Miani nr.2, 33047 Remanzacco, Italy
         \and
              Istituto Spezzino Ricerche Astronomiche, 
              via Castellazzo 8/d 19123 La Spezia, Italy
         \and
              Stazione Astronomica di Sozzago, 
              28060 Sozzago, Italy 
             }

   \date{Received 03 October 2007 ; accepted 18 October 2007}

\abstract { 
We report the detection of transits by the $3.1 M_{\rm Jup}$ 
companion to the V=8.17 G0V star \object{HD 17156}. The transit was 
observed by three independant observers on Sept. 9/10, 2007 (two in 
central Italy and one in the Canary Islands), who obtained detections at 
confidence levels of 3.0~$\sigma$, 5.3~$\sigma$, and 7.9~$\sigma$, 
respectively. The observations were carried out under the auspices of 
the Transitsearch.org network, which organizes follow-up photometric 
transit searches of known planet-bearing stars during the time intervals 
when transits are expected to possibly occur. Analyses of the 
7.9~$\sigma$ data set indicates a transit depth $d=0.0062 \pm 0.0004$, 
and a transit duration $t=186 \pm 5$ min. These values are 
consistent with the transit of a Jupiter-sized planet with an impact 
parameter $b=a \cos{i}/R_{\star} \sim 0.8$. This planet occupies a 
unique regime among known transiting extrasolar planets, both as a 
result of its large orbital eccentricity ($e=0.67$) and long orbital 
period ($P=21.2 {\rm d}$). The planet receives a 26-fold variation in 
insolation during the course of its orbit, which will make it a useful 
object for characterization of exoplanetary atmospheric dynamics.
}


\maketitle

\section{Introduction} 

During the past several years, the discovery rate of transiting planets 
has begun to increase rapidly, and twenty transiting planets with secure 
characterizations are currently known \footnote{ 
Extrasolar Planets Encyclopedia {\tt http://exoplanet.eu}}.
This aggregate consists mostly of short-period hot-Jupiter type planets, 
with prototypical examples being HD~209458b 
\citep{2000ApJ...529L..45C,2000ApJ...529L..41H} and HD~189733b 
\citep{2005A&A...444L..15B}. These planets tend to have $M\sim 1M_{\rm 
Jup}$, $2 {\rm d}<P<5{\rm d}$, and tidally circularized orbits.

In the past year, two remarkable discoveries have significantly extended 
the parameter space occupied by known transiting planets. HD~147506b 
\citep{2007arXiv0705.0126B} with $M=8.04 \, M_{\rm Jup}$ is by far the most 
massive planet known to exhibit transits. It also has the longest 
orbital period (5.63 days) and a startlingly large orbital eccentricity, 
$e\sim 0.5$. At the other end of the mass scale, Gl~436b 
\citep{2004ApJ...617..580B,2007A&A...472L..13G} has $M=0.07 M_{\rm 
Jup}$, a 2.64 day orbital period, and an eccentricity $e=0.15\pm0.01$ 
\citep{2007arXiv0707.2778D}. These two planets straddle more than a 
hundred-fold difference in mass, and their significant 
non-zero eccentricities are also capable of imparting 
important information.

At present, infrared observations of transiting extrasolar 
planets by Spitzer present an incomplete and somewhat contradictory 
overall picture. It is not understood how the wind vectors and 
temperature distributions on the observed planets behave as a function 
of pressure depth, and planetary longitude and latitude. Most 
importantly, the effective radiative time constant in the atmospheres of 
short-period planets remains unmeasured, and as a result, dynamical 
calculations of the expected planet-wide flow patterns 
\citep{2003ApJ...587L.117C,2005ApJ...629L..45C,2005ApJ...618..512B,2007ApJ...657L.113L,2007arXiv0704.3269D} 
have come to no consensus regarding how the surface flow should appear. This 
lack of agreement between the models stems in large part from the 
paucity of unambiguous measurements of the radiative time constant in 
the atmosphere. What is needed, is a transiting planet with both a 
long-period orbit and a large orbital eccentricity. If such a planet 
were known, then one could use Spitzer to obtain infrared time-series 
photometry of the planet during the periastron passage. The transit 
guarantees knowledge of both the geometric phase function and the 
planetary mass. This information would in turn allow a measured rate of 
increase in flux to inform us of the planet's atmospheric radiative time 
constant in the observed wavelength regime.

The orbital periods of the known transiting planets are all 
significantly shorter than 6 days. This bias is due both to the 
intrinsically lower geometric probability of transit as one moves to 
longer periods, and also to the fact that ground-based wide-field 
transit surveys that rely on photometry folding become very 
significantly incomplete for planets with orbital periods longer than 5 
days. If one wants to detect longer-period transiting planets from the 
ground, a more productive strategy is to monitor known RV-detected 
planet-bearing stars at the times when the radial velocity solution 
suggests that transits may occur. This strategy has the further 
advantage of producing transits around stars that tend to be both bright 
and well-suited for follow-up observations.

Long-period transiting planets present an ideal observing opportunity 
for small telescope observers. \cite{2003PASP..115.1355S} have 
suggested that a global network of telescopes, all capable of 
$\sim1\%$ photometry can easily outperform a single large telescope in 
terms of efficiency of transit recovery. Since inception in 2002, the 
Transitsearch.org network has conducted follow-up 
searches on a number of intermediate-period planets
(see e.g. \citealt{2006ApJ...653..700S}). 

The Doppler-based discovery of HD~17156b was recently published by the N2K consortium 
\citep{2007arXiv0704.1191F}. The planet has
M$\sin i = 3.12{\rm M_{Jup}}$, with $P=21.22$ days and 
$e\sim0.67$. 
\cite{2007arXiv0704.1191F} report that the V=8.17 G0V host star has
$M= 1.2 M_\odot$ and $R= 1.47 R_\odot$. The planet's 
semi-major axis $a=0.15 {\rm AU}$ thus indicates a periastron distance 
of $a_{min}=0.0495 {\rm AU}=7.2 R_{\star}$. A best fit to the radial 
velocities indicates longitude of periastron $\omega=121 \pm 
11^{\circ}$. The orbital orientation is favorable,
yielding an a-priori geometric transit probability of
$P\sim13$\%.

In their discovery paper, \cite{2007arXiv0704.1191F} reported 241 
individual photometric measurements obtained over a 179 day
interval, and with a mean dispersion $\sigma=0.0024$ mag. No 
significant rotation-induced periodicity was seen. Together, the 
observations sampled approximately 25\% of the $1-\sigma$ transit 
window, and no evidence for a transit was observed.
After the \cite{2007arXiv0704.1191F} discovery paper was made public, 
the star was added to the Transitsearch.org candidates list\footnote 
{http://207.111.201.70/transitsearch/dynamiccontent/candidates.html} and 
observers throughout the Northern Hemisphere were repeatedly encouraged 
to obtain photometry of the star\footnote{see www.oklo.org}. The first 
available window of opportunity occurred on 9/10 Sept., 2007, with 
the transit midpoint predicted to occur at HJD $2454353.65 \pm 0.30$. 

\section{Observations}

We collected data from different observatories during the night of 
September 9/10, 2007. The following instrumentation was used:
\begin{itemize}
\item {\it Almenara:}
Observations were gathered in R band and 7 seconds of exposure time with 
the TELAST 0.30 m telescope, a stellar photometer devoted to IAC 
Asteroseismology programs. The telescope is a {\it f}/10 
Schmidt-Cassegrain catadioptric, an SBIG STL-1001E CCD camera provides a 
field of view of 29$'\times29'$ (scale 1.7\arcsec/px). The night was 
windy, affecting the telescope (totally exposed), the stars appeared doubled 
and even tripled due to this. The gap in the center of the observations 
was caused by the lost of the guide star due to the wind.
\item{\it Bissinger:}
Observations were made from Pleasanton, California USA 
using a 0.4 m diameter modified Schmidt-Cassegrain 
telescope operating at {\it f}/6 with an SBIG ST-10XME CCD camera and a 
Bessell I band filter.  Imaging began at 04:05 UT on 10 Sept.
and ended at 09:09 UT on 10 Sept. with an exposure cadence of 
43 seconds. Bins of 15 exposures were made producing a flat light curve 
with an r.m.s. of 0.003 mag.
\item{\it Gasparri:}
The telescope used is a commercial 0.25 m {\it f}/4.8 Newtonian, located 
near Perugia, Italy. The camera is a SBIG ST-7XME with KAF-0402 CCD 
providing a field of view of 19.8$'\times13.2'$ with sampling of 
1.55\arcsec/px. The photometric observation started at 20 UT on 
9 Sept. and stopped at 04 UT on Sept. 10. The presence of some 
clouds and veils limit the useful data to 00-02 UT of 10 Sept.
Exposures were made through a near-IR filter, and are of 20s duration, with
339 useful images collected.
\item{\it Lopresti:}
Observations were conducted at La Spezia, Italy with a Maksutov--Newton 
telescope of 0.18 m diameter {\it f}/4 and an SBIG ST-10xme CCD camera, 
with a framed field of $70'\times47'$ (scale 1.7\arcsec/px). Observations
run from 22 UT 9 Sept. through 04 UT 10 Sept. Exposure times were 5s.
A total of 580 R band images were collected.
\item{\it Manzini:}
The Stazione Astronomica di Sozzago is an observatory located in Sozzago 
(Novara) Italy (international code IAU A12). Observations were conducted
with a 0.40 m Cassegrain telescope {\it f}/6.7, equipped with a CCD camera
Hi-SIS43ME (FoV $18'\times11'$, scale 0.7\arcsec/px). 
Useful data were collected until 20 UT 9 Sept., when clouds intervened.
\end{itemize}

\section{Data Analysis}
All the raw images were calibrated in the standard way; each 
observer took a series of images to correct for the irregular pixel sensivity
(flat-fielding) and dark current effects. Out of the 6 
datasets, only three cover the central transit window. These were obtained
by the amateur astronomers Lopresti and Gasparri (in Italy), and by
Almenara (at the IAC). Unfortunately, the data are irregular 
in their coverage of the transit and in their precision.

The three data sets were analyzed with IDL
routines to perform simple aperture photometry. The center of the 
aperture was calculated by a gaussian fit, and the aperture was held
fixed (to 15-20 px in radius, depending on the data set). We 
removed the sky background contribution after an estimation of its value 
in an annulus around the target aperture. The brightest stars in the 
field were measured the same way, and a reference light curve was constructed 
by adding the flux of these stars. The target flux was divided by this 
reference to get the final normalized curve. 
The data included in this detailed analysis are: 
\begin{itemize} 
\item Almenara (A): From 6 hours before center to 3.4 hours after 
transit center. The dispersion is high at the beginning, and better at 
the end, decreasing from 0.5\% to 0.4\% (in a k-sigma filtered version 
of the original light curve).  Two stars were used to 
build the reference light curve.
\item Gasparri (G): From 2 hours before center to 1.1 hours after 
center. Three stars were used to build the reference star. The r.m.s. of 
the residuals is $\sim$0.4\%. Soon before transit center there is a rise 
in the light curve that we were not able to correct, due to the clouds 
and veils affecting differently the target and each of the stars used to 
build the reference.
\item Lopresti (L): From 4 hours before transit center to 1.5 hours 
after transit center. The last 45 minutes were taken with a 180$^\circ$ 
rotation of the CCD (due to the mount configuration). We have not been 
able to fully correct for the effect of this rotation, as data are sensitive
to uncorrected minor flat field effects. To avoid the 
introduction of offsets, we have thus ignored these last data. Three 
stars were used to build the reference light curve. 
The r.m.s. of the residuals is $\sim$0.9\%.
\end{itemize}
\begin{figure}%[htbp]
   \resizebox{\hsize}{!}{
   \includegraphics[angle=-90]{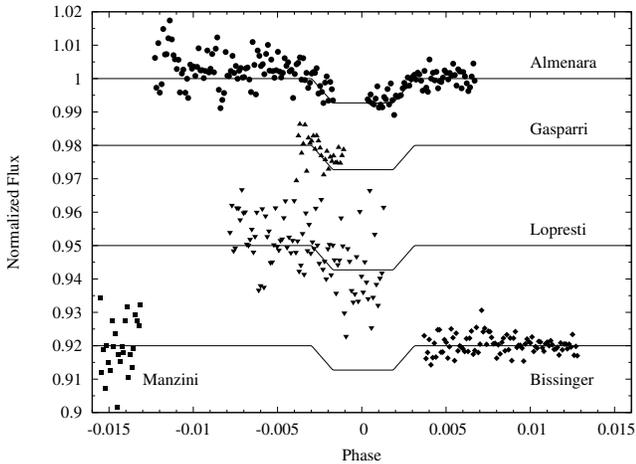}}
   \caption{Normalized light curves of HD~17156 at the moment of transit, 
   and the best fitted trapezoid to the Almenara data set (solid lines). 
   Data sets of different observers have been shifted each other for clarity. 
   The transit is centered in HJD~2454353.61}
   \label{figlc}
\end{figure}
The final data sets are plotted in the Fig.\ref{figlc}.
We performed a fit to a trapezoidal function, with four free parameters: 
center, width, depth and size of the transition between the two levels 
of the trapezoid. The significance of the transit detection was then 
evaluated as the value of the depth of the trapezoid divided by the 
dispersion of the fit. Thus, we obtain a 5.6~$\sigma$ detection for the 
(A) set. This value increases to 7.9~$\sigma$ once the baseline is 
corrected by a parabolic fit to the parts of the light curve outside of 
the trapezoid, plus an extra margin of 0.001 in phase to avoid the 
inclusion of points inside transit. For the (L) and (G) data sets the 
free parameters for the trapezoid fitting were only the depth and 
center. The other two parameters were set to the result of the fit in 
the (A) set. We obtain the values of 5.3~$\sigma$ and 2.97~$\sigma$, 
respectively.

Due to the quality of the three data sets, we believe it is too
dangerous to perform a combined analysis; 
specially the zero offsets are not too clear in the Gasparri (G) and 
Lopresti (L) data sets, and they might dominate the result of a fitting 
to a combined light curve. We thus analyzed the
most homogeneous and least noisy part of the data, namely the 
egress recorded in the (A) set, to determine the main characteristics of 
the transit. We employed two strategies: fitting to a trapezoid to 
estimate the depth of the transit, and fitting to a model of an 
eccentric transit, following the formalism of 
\cite{2006A&A...450.1231G}.

In order to correct for the baseline,
the trapezoid fit to the egress was performed in two steps:
(1) A first trapezoid is fitted, removed from 
the light curve, and a line is fitted to the residuals. This line is 
removed from the original light curve and (2) a second trapezoid is 
fitted, providing the values of the depth and time of egress. The errors 
are evaluated by a bootstrap analysis, performing 20\,000 tests with 
data sets in which 50\% of the residuals points were randomly re-sorted, 
and the best fitted model was re-added to the residuals to build the 
data set. The same two-step fitting was performed in each data set. The 
depth of the trapezoid was found to be 0.0062$\pm$0.0004.

The second strategy consisted in a fit to the equations of 
\cite{2006A&A...450.1231G}. The fitted parameters were the phase of 
start of the transit, $k$, $i$, and three coefficients defining a 
parabolic baseline correction. The two non-linear limb darkening 
coefficients were fixed to
u$_+=0.65$ and u$_-=-0.05$ (see \cite{2006A&A...450.1231G} for a 
definition of these coefficients) from the tables of 
\cite{2000A&A...363.1081C} for ATLAS stellar models. The eccentricity 
and the longitude of the periastron were fixed to values obtained from 
\cite{2007arXiv0704.1191F}. The best solution was obtained by minimizing 
the $\chi^2$ between the model and the observations using the algorithm 
AMOEBA \citep{1992nrfa.book.....P}. The errors were estimated by a 
bootstrap analysis similar at that described above, performing 1\,500 
tests. The best fitted values using this technique are reported in 
Table\ref{tab1}.

\begin{figure}%[htbp]
   \resizebox{\hsize}{!}{
   \includegraphics{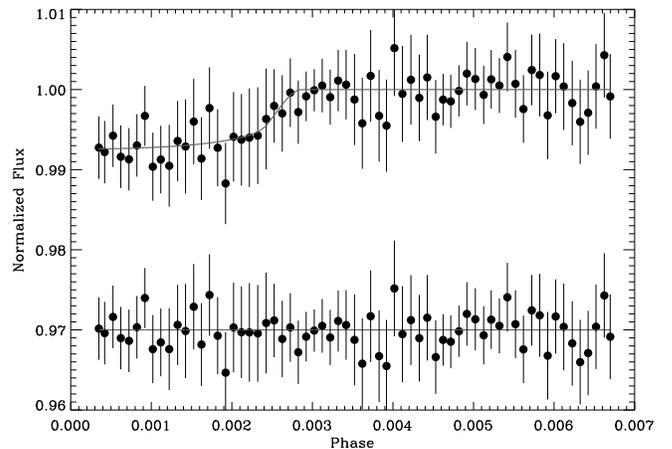}}
\caption{Top: Normalized phase plot of the egress of the transit of 
HD~17156 from the Almenara data set. The error bar in each point it was 
calculated as the standard deviation of the 20 points closest to the 
point i. The plotted error bars are 2 times this quantity. The 
overplotted line is the best fitted model using the formalism of Gimenez 
as described in the text. Bottom: the residuals of the fit.}
   \label{gimenez}
\end{figure}

Based on the inclination, the planet mass is then $M_p = 3.12  \pm 0.5 
M_{\rm Jup}$. Based on the radius of the star and the $k$ determination, 
the radius of the planet is $R_p = 1.15  \pm 0.11  R_{\rm Jup}$, 
and the resulting planet mean density is $\rho_p = 2.58 \pm 0.84$ g/cm$^3$.
These properties are summarized in Table\ref{tab1}, the final fit and the 
residuals are plotted in Fig.\ref{gimenez}.

\begin{table}
   \caption{Transit fit and planetary parameters for HD~17156 b}
   \label{tab1}
   \centering
   \begin{tabular}{l r}
   \hline\hline
   Parameter                      &    Value\\
   \hline
   $T_{mid}$ (HJD)                &    2\,454,353.61 $\pm$ 0.02     \\
   $\phi_{\rm  egress}$           &          0.003050$\pm$ 0.000075  \\
   Transit duration (day)         &          0.1294  $\pm$ 0.0367    \\
   $k=R_p/R_\star$                &          0.08007 $\pm$ 0.0028    \\
   $i$ (deg)                      &            87.89 $\pm$ 0.10      \\
   \hline
   $R_p$ ($R_{\rm Jup}$)          &         1.15     $\pm$ 0.11      \\
   $\rho_p$ (g/cm$^3$)            &         2.58     $\pm$ 0.84      \\
   \hline
   \end{tabular}
\end{table}

As additional test we have checked the Hipparcos photometry 
\citep{1997ESASP1200.....P}. Hipparcos observed HD~17156 (HIP~13192) on 
142 occasions with a standard deviation of 0.0013 mag. An inspection of the
light curve folded with the orbital period of the planet shows only
two photometric points close to the transit window.

During the night of September 30 / October 1, HD~17156b was observed from the 
Mount Laguna Observatory in southern California. 
The team composed by William Welsh, Abhijith Rajan, Jonathan Irwin,
Philip Nutzman and David Charbonneau kindly report to us
(Charbonneau, personal communication)
that the transit ingress was observed near Oct 1 UT 6:30, and
a flat bottomed event followed, egress was lost due to clouds.

In addition, observer Davies, located
in Torrance, California, obtained $\sim10,000$ CCD images which 
cover the full transit window. A preliminary analysis confirms the detection
of the transit, and a more detailed analysis will be presented
in a forthcoming paper (Irwin et al., in preparation).

\section{Discussion}

The detection of transits by a planet with a three-week orbital period
demonstrates the utility of ad-hoc networks of 
small telescopes for obtaining photometric follow-up of planets whose 
orbital parameters have been determined via Doppler radial velocities.
Indeed, the transits of HD~17156b offer a
plethora of interesting opportunities for follow-up 
observations.

With its high orbital eccentricity and small periastron distance, 
HD~17156b appears to bear a curious kinship to 
HD~80606b, HD~147506b, and HD~108147b. All 
three of these planets occupy a locus of the $a$--$e$ plane where they 
should actively be undergoing tidal dissipation, and therefore they 
should be generating significant quantities of excess interior heat. 
Our measurement indicates that tidal heating is not 
significantly inflating the planetary radius. 
The nominal $R=1.1 R_{\rm Jup}$ radius 
predicted by baseline models (e.g. those of \citealt{2003ApJ...592..555B})
is confirmed by our observations.

Follow-up photometric measurements during future transits will allow a 
more accurate determination of the orbital inclination of HD~17156b. An 
improved value for $i$, in turn, will generate an accurate assessment of 
likelihood that the planet can be observed by Spitzer in secondary 
transit, and will enable a much-improved constraint on the 
still-uncertain radius of the parent star. In the event that secondary 
transits can be observed, a direct measurement of the excess tidally 
generated luminosity from the planet is a distinct possibility (see e.g. 
\citealt{2007arXiv0707.2778D}).

As a consequence of its highly eccentric orbit, HD~17156b experiences a 
26-fold variation in insolation during the 10.6 day interval between 
periastron and apoastron. This extreme radiative forcing may drive
interesting, and potentially observable dynamical atmospheric 
flows on the planet \citep{2007ApJ...657L.113L}. The large 
tidal forces experienced during periastron have almost certainly forced 
the planet into pseudo-synchronous rotation (e.g. 
\citealt{1968ARA&A...6..287G,1981A&A....99..126H,2005MNRAS.364L..66P}). 
Rotationally induced 
modulation in the infrared light curve following periastron is 
potentially observable, and may be of great utility in selecting between 
the current divergent predictions for the actual value of the 
pseudo-synchronous spin frequency.

HD~17156b is quite massive, as is often the case for planets
orbiting one member of a binary pair \citep{2007A&A...462..345D},
and the eccentricity is large.
These characteristics favor a formation scenario involving migration
and/or dynamical evolution
in the presence of a sufficiently close external perturber.
Such a perturber could be either a companion star or an additional planet(s) in the system.

For some of the close-in planets with $m \sin i > 1.5~M_{\rm Jup}$ 
orbiting single stars, there are already indications of a history of
significant dynamical perturbations. For example, HD~118203b, HD~68988b 
and HIP~14810b all have anomalously high eccentricities
that may be indicative of additional 
perturbing companions, perhaps with masses below (or periods longer than) the threshold of 
immediate radial velocity detection (this is certainly the case for HD~68988b and 
HIP~14810b, which both have long-period planetary companions 
\citealt{2006ApJ...646..505B}).
Due to mutual perturbations, the eccentricities of the bodies in the precursor system may
have grown to the point where crossing orbits were achieved.
Repeated close encounters among the planets would have then
generated a period of chaotic evolution that typically terminates
with the ejection of one planet on a hyperbolic trajectory \citep{2002Icar..156..570M}.

Alternately, a stellar companion
could also effectively trigger dynamical evolution or
instability in a precursor system, eventually leading to the  
current configuration (for some examples, see
\citealt{2007A&A...467..347M,2007A&A...472..643M,2003ApJ...589..605W}).
With reference to a stellar companion,
a quick inspection of POSSI, POSSII, and 2MASS images do not reveal
any clear association between faint field stars and HD~17156.
The only potentially interesting source is 2MASS~02494068+7144583.
It shows an appreciable proper motion, but in the opposite direction
of the proper motion of HD~17156. The star lies 22.2\arcsec~from 
HD~17156; if they are at the same distance, the apparent
separation is $\sim$1740 AU. 

A combination of continued radial 
velocity monitoring of HD~17156, in conjunction with accurate 
measurements of successive transit midpoints, gives hope for the 
detection and accurate characterization of additional bodies in the 
system via a novel set of constraints.

\begin{acknowledgements}
Support for M.B. is provided by Agence Nationale de la Recherche (France).
Support for R.A. is provided by CNES under contract 02/0425.
Support for Transitsearch.org was provided by NASA under grant NNG04GN30G 
through the OSS program. 
We wish to thank financial support from spanish grants ESP2004-03855-C03 and
AYA2004-04462.
We thank Dr. David Charbonneau for generous advice.
M.~B. thanks C. Lopresti for useful discussions, and R. Calanca 
is gratefully aknowledged for his invaluable support in the divulgation 
the exoplanetary science between the italian amateur astronomical community.
\end{acknowledgements}

\end{document}